\documentclass[groupedaddress,aps,pra,superscriptaddress]{revtex4}%
\usepackage{epsfig,dsfont,amssymb,amsmath,amsthm,amsfonts,amsbsy,mathrsfs}
\usepackage{graphicx}
\usepackage{amsmath}
\usepackage{amssymb}
\usepackage{mathdots}
\usepackage{float}
\begin{document}
\baselineskip18pt
\title{Super quantum discord for X-type states}
\author{Tao Li}
\affiliation{School of Mathematical Sciences,  Capital Normal University,  Beijing 100048,  China}
\author{Teng Ma}
\affiliation{School of Mathematical Sciences,  Capital Normal University,  Beijing 100048,  China}
\author{Yaokun Wang}
\affiliation{School of Mathematical Sciences,  Capital Normal
University,  Beijing 100048,  China}
\affiliation{Department of Mathematics,  Tonghua Normal University, Tonghua 134001,  China}
\author{Shaoming Fei}
\affiliation{School of Mathematical Sciences, Capital Normal University, Beijing 100048, China}
\affiliation{Max-Planck-Institute for Mathematics in the Sciences, 04103 Leipzig, Germany}
\author{Zhixi Wang}
\affiliation{School of Mathematical Sciences,  Capital Normal
University,  Beijing 100048,  China}


\begin{abstract}
\baselineskip15pt

Weak measurement is a new way to manipulate and control quantum systems. Different from projection measurement, weak measurement only makes a small change in status. Applying weak measurement to quantum discord, Singh and Pati proposed a new kind of quantum correlations called ``super quantum discord (SQD)" [Annals of Physics \textbf{343},141(2014)]. Unfortunately, the super quantum discord is also difficult to calculate. There are only few explicit formulae about SQD.  We derive an analytical formulae of SQD for general X-type two-qubit states, which surpass the conclusion for Werner states and Bell diagonal states. Furthermore, our results reveal more knowledge about the new insight of quantum correlation and give a new way to compare SQD with normal quantum discord. Finally, we analyze its dynamics under nondissipative channels.
\end{abstract}
\maketitle

\section {Introduction}

The quantum entanglement plays important roles in quantum
information processing \cite{nielsen}. However, besides quantum
entanglement there are other quantum correlations also useful for quantum information processing. It is found that many
tasks can be carried out with quantum correlations other than
entanglement \cite{horodecki2,bennett,niset}. In particular, the
quantum discord
\cite{ollivier,bylicka,werlang,sarandy,ferraro,fanchini,dakic,modi,luo1,luo,libo,lang,ali,Mazzola,Maziero}
plays an important role in some quantum information processing like to
assist optimal state discrimination, in which only one side
discord is required in the optimization process of assisted state
discrimination, while another side discord and entanglement is not
necessary \cite{ost}.

Unfortunately, underlying quantum measurements process quantum states are fragile. When we measure a quantum state in some orthogonal basis, since quantum discord is defined by the POVM quantum measurement, the coherence
of the state has been loosened. Taking account of quantum states' potential privacy, it is reasonable to find a solution to deal with this problem. Such a solution was making use of weak measurement which induced by Aharonov-Albert-Vaidman\cite{Aharonov}. Applying such a scheme, we can replace the POVM measurement by weak measurement in the definition of quantum discord, which gives rise to so called super quantum discord (SQD) \cite{Pati}.

Super quantum discord sheds a new insight on the nature of quantum correlation. It also has vivid properties, such as the monotony. Super quantum discord not only a new insight in fundamental physics but also useful in applications. But super quantum discord is difficult to calculate. There are only few explicit formulae about SQD. The analytical formulae of Werner states \cite{Pati} and Bell diagonal states \cite{Wang} are only two results. In order to obtain more useful results,  we take a new method to compute more general states---X-type states, which including Werner states and Bell diagonal states. Obviously, our results include the results in \cite{Pati} and \cite{Wang}. Furthermore, in order to show the potential property of technological implications, we analyze the dynamics of two-qubit X-states under non-dissipative channels and compare super quantum discord with discord by using the explicit formulae. This is exemplified by the fact that the super quantum discord often larger than the quantum discord defined by projective measurement. Also, the super quantum discord sometime captures more quantum correlations.

This paper is organized as follows. In section \uppercase\expandafter{\romannumeral2}, we derive analytical formulae of super quantum discord for X-states. In section \uppercase\expandafter{\romannumeral3},  we compute the  super quantum discord of some concrete examples, and analyze their dynamics under nondissipative channels, we further compare it with discord and present some new property of super quantum discord.

\section {Super quantum discord for two-qubit X-states}

Super quantum discord of some special states has been computed recently, the Werner states and the Bell diagonal states are included. Now, we extend the results about super quantum discord in \cite{Pati} and \cite{Wang} to the whole two-qubit X-states. Let us consider a two-qubit $X$-state:
\begin{align}
\rho_{X}=\left(\begin{array}{cccc}a_{11} & 0 & 0 & a_{14}\\0& a_{22} & a_{23} & 0 \\
0 & a^*_{23} & a_{33} & 0\\a^*_{14} & 0 & 0 & a_{44}\end{array}\right),
\end{align}
where $\sum\limits_{i=1}\limits^4a_{ii}=1$, $|a_{23}^2|\leq
a_{22}a_{33}$, $|a_{14}^2|\leq a_{11}a_{44}$.
The density matrix $\rho_{X}$ can be written
as \cite{Shi}:
\begin{align}
\rho_{X}=\frac{1}{4}\left(\begin{array}{cccc}1+d_{1} & 0 & 0 & c_{1}-c_{2}\\0& 1+d_{2} & c_{1}+c_{2}
& 0 \\0 & c_{1}^*+c_{2}^* & 1+d_{3} & 0\\c_{1}^*-c_{2}^* & 0 & 0 & 1+d_{4}\end{array}\right),
\end{align}
where  $c_{1}$ and $c_{2}$ are complex, $d_1$, $d_2$, $d_3$ and
$d_4$ are real, $d_{1}=c_{3}+a_{3}+b_{3}$,
$d_{2}=-c_{3}+a_{3}-b_{3}$, $d_{3}=-c_{3}-a_{3}+b_{3}$,
$d_{4}=c_{3}-a_{3}-b_{3}$. These parameters are determined by the
entries of the density matrix, $a_{3}=a_{11}-a_{44}+a_{22}-a_{33}$,
$b_{3}=a_{11}-a_{44}-a_{22}+a_{33}$,
$c_{3}=a_{11}+a_{44}-a_{22}-a_{33}$, $c_{1}=2(a_{23}+a_{14})$,
$c_{2}=2(a_{23}-a_{14})$.

Let $\{\Pi^B_i\}$, $i=0,1$, be the projective measurements. The
discord of a bipartite quantum state $\rho_{AB}$ with the
measurement $\{\Pi^B_i\}$ on the subsystem $B$ is the dissimilarity
between the mutual information $I$($\rho_{AB}$) \cite{Partovi} and the
classical correlation $J_{B}(\rho_{AB}$) \cite{Henderson}:
\begin{align}
D(\rho_{AB})=\min\limits_{\{\Pi_i^B\}}\sum\limits_ip_iS(\rho_{A|i})+S(\rho_B)-S(\rho_{AB}),
\end{align}
where the minimization goes over all projective measurements
\{$\Pi_i^B$\}, $S(\rho)=-{\rm tr}(\rho\log_2\rho)$ is the von Neumann
entropy of a quantum state $\rho$,  $\rho_B$ is the reduced density
matrices of $\rho_{AB}$ and
\begin{align}
p_i={\rm tr}_{AB}[(I_A\otimes\Pi_i^B)\rho_{AB}(I_A\otimes\Pi_i^B)],\\
\rho_{A|i}=\frac{1}{p_i}{\rm tr}_{B}[(I_A\otimes\Pi_i^B)\rho_{AB}(I_A\otimes\Pi_i^B)].
\end{align}

The weak measurement operators are given by \cite{Oreshkov},
\begin{align}
P(+x)=\sqrt{\frac{1-\tanh x}{2}}\Pi_0+\sqrt{\frac{1+\tanh x}{2}}\Pi_1,
\\P(-x)=\sqrt{\frac{1+\tanh x}{2}}\Pi_0+\sqrt{\frac{1-\tanh x}{2}}\Pi_1,
\end{align}
where $\Pi_0$ and $\Pi_1$ are two orthogonal projectors satisfying
$\Pi_0+\Pi_1=I$, $x$ is the strength parameter of measurement. The
weak measurement operators satisfy: (\romannumeral1)
$P^\dagger(+x)P(+x)+P^\dagger(-x)P(-x)=I$, (\romannumeral2)
$\lim\limits_{x \to \infty}P(+x)=\Pi_0$ and $\lim\limits_{x \to
\infty}P(-x)=\Pi_1$.

The super quantum discord is defined by \cite{Pati}:
\begin{align}
D_w(\rho_{AB})=\min\limits_{\{\Pi_i^B\}}S_w(A|\{P^B(x)\})+S(\rho_B)-S(\rho_{AB})\nonumber,
\end{align}
where
\begin{align}
S_w(A|\{P^B(x)\})=p(+x)S(\rho_{A|P^B(+x)})+p(-x)S(\rho_{A|P^B(-x)}),\\\nonumber
p(\pm x)={\rm tr}_{AB}[(I_A\otimes P^B(\pm x))\rho_{AB}(I_A\otimes P^B(\pm
x))],\nonumber \\[2mm]\nonumber
\rho_{A|P^B(\pm
x)}=\frac{{\rm tr}_{B}[(I_A\otimes P^B(\pm x))\rho_{AB}(I_A\otimes P^B(\pm
x))]}{{\rm tr}_{AB}[(I_A\otimes P^B(\pm x))\rho_{AB}(I_A\otimes P^B(\pm
x))]},\nonumber
\end{align}
where $\{P^B(x)\}$ is the weak measurement operators on subsystem
$B$.

The weak measurement operators on subsystem $B$ can be expressed as
\begin{align}
I_A\otimes P^B(\pm x)=\sqrt{\frac{1\mp\tanh x}{2}}I\otimes V\Pi_0V^\dagger+\sqrt{\frac{1\pm\tanh x}{2}}I\otimes V\Pi_1V^\dagger,
\end{align}
where $\Pi_k=|k\rangle\langle k|$, $k=0,1$, $|k\rangle$ is the
computational base, and $V$ is a $2\times2$ unitary transformation. $V$ can be generally
expressed as \cite{luo}:
\begin{align}
V=tI+i\textbf{y}\cdot\textbf{$\sigma$},
\end{align}
where $ \textbf{y}=(y_1,y_2,y_3)$ and $t,y_1,y_2,y_3\in R^1$, $t^2+y_1^2+y_2^2+y_3^2=1$.

To evaluate the super quantum discord of $\rho_X$, let us first
express $\rho_X$ in terms of the bases $I\otimes
I,\sigma_i\otimes\sigma_j$, $i,j=0,1,2$.
\begin{eqnarray}
&& \rho_{X}=\frac{1}{4}(I+\sum_i\Re(c_i)\sigma_i\otimes\sigma_i)
+\frac{1}{4}[(b_3-a_3)I\otimes\sigma_3+(\Im(c_1)+\Im(c_2)\sigma_1\otimes\sigma_2)],\nonumber
\end{eqnarray}
where $\Re(c_i),\Im(c_i)$ are the real and complex parts of $c_i$.
By using the relations
\begin{eqnarray*}
&&V^\dagger\sigma_1V=(t^2+y^2_1-y_2^2-y_3^2)\sigma_1+2(ty_3+y_1y_2)\sigma_2
+2(-ty_2+y_1y_3)\sigma_3,\\\nonumber
&&V^\dagger\sigma_2V=(t^2+y^2_2-y_1^2-y_3^2)\sigma_2+2(ty_1+y_2y_3)\sigma_3
+2(-ty_3+y_1y_2)\sigma_1,\\\nonumber
&&V^\dagger\sigma_3V=(t^2+y^2_3-y_1^2-y_2^2)\sigma_3+2(ty_2+y_1y_3)\sigma_1
+2(-ty_1+y_2y_3)\sigma_2,\nonumber
\end{eqnarray*}
$\Pi_0\sigma_3\Pi_0=\Pi_0$, $\Pi_1\sigma_3\Pi_1=-\Pi_1$,
$\Pi_j\sigma_k\Pi_j=0$ for $j=0,1$, $k=1,2$ in \cite{luo}. Setting
$a_1=z_1\Re(c_1)+z_2\Im(c_2)$, $a_2=z_2\Re(c_2)-z_1\Im(c_1)$, with
$z_1=2(-ty_2+y_1y_3)$, $z_2=2(ty_1+y_2y_3)$,
$z_3=t^2+y^2_3-y^2_1-y_2^2$, we have the ensemble $\{\rho_{A|P^B(\pm x)},\,p(\pm x)\}$ after weak measurement, from Eqs.(4) and
(5)
\begin{align}
&p(+x)=\frac{1}{2}(1-b_3z_3\tanh x),~~~~~~\rho_{A|P^B(+x)}=\frac{1}{2}\left[I+\frac{\tanh x(-a_1\sigma_1-a_2\sigma_2)+(a_3-c_3z_3\tanh x)\sigma_3)}{1-b_3z_3\tanh x}\right],\\
&p(-x)=\frac{1}{2}(1+b_3z_3\tanh x),~~~~~~\rho_{A|P^B(-x)}=\frac{1}{2}\left[I+\frac{\tanh x(a_1\sigma_1+a_2\sigma_2)+(a_3+c_3z_3\tanh x)\sigma_3)}{1+b_3z_3\tanh x}\right].
\end{align}
The eigenvalues of $\rho_{A|P^B(+x)}$ and $\rho_{A|P^B(-x)}$ are given by
\begin{align}
& \frac{1}{2}\left[1\pm \frac{\sqrt{(a_3-c_3z_3\tanh x)^2+(a_2^2+a_1^2)\tanh^2 x}}{1-b_3z_3\tanh x}\right],\\
& \frac{1}{2}\left[1\pm \frac{\sqrt{(a_3+c_3z_3\tanh x)^2+(a_2^2+a_1^2)\tanh^2 x}}{1+b_3z_3\tanh x}\right].
\end{align}

We now compute the minimum value of $S(\rho_{A|P^B(+x)})$ and the
corresponding $p(+x)$ by using the method of Hessian matrix and the symmetries in Eqs.(9)-(13). In order to avoid redundant narrating, we only give the result in the following tables, and the minimum value lies at $z_3=0$ or $z_3=1$. The extremum lies at the following points:
\begin{align*}
\renewcommand\arraystretch{1.6}
\large
\begin{tabular}{|c|c|c|}
\multicolumn{3}{c}{Table \uppercase\expandafter{\romannumeral1}: The minimum value of $S(\rho_{A|P^B(+x)})$ and
$p(+x)$}\\
\hline
($z_3,z_2,z_1$)& $p(+x)$& $\lambda_{\pm}$\\
\hline
(1,0,0) & $\frac{1-b_3\tanh x}{2}$ & $\frac{1}{2}\scriptstyle \left[1\pm\frac{a_3-c_3\tanh x}{1-b_3\tanh x}\right]$ \\
\hline
(0,0,1)& $\frac{1}{2}$& $\frac{1}{2}\scriptstyle \left[1\pm+|a_3|\cdot\tanh x \right]$\\
\hline
(0,1,0)& $\frac{1}{2}$& ${\scriptstyle\frac{1}{2} \left[1\pm\sqrt{a_3^2+(|c_2|^2-|c_1-c_2|^2-|c_1+c_2|^2+(\Re(c_1-c_2)+\Im(c_1+c_2))^2+(\Re(c_1+c_2)+\Im(c_1-c_2))^2)\tanh^2 x }\right]}$\\
\hline
(0,-1,0)& $\frac{1}{2}$& ${\scriptstyle\frac{1}{2} \left[1\pm\sqrt{a_3^2+(|c_2|^2-|c_1-c_2|^2-|c_1+c_2|^2+(\Re(c_1-c_2)-\Im(c_1+c_2))^2+(\Re(c_1+c_2)-\Im(c_1-c_2))^2)\tanh^2 x }\right]}$\\
\hline
\end{tabular}
\end{align*}
Similarly for the minimum value of $S(\rho_{A|P^B(-x)})$ and
$p(-x)$, we have:
\begin{align*}
\renewcommand\arraystretch{1.6}
\large
\begin{tabular}{|c|c|c|}
\multicolumn{3}{c}{Table \uppercase\expandafter{\romannumeral2}: The minimum value of $S(\rho_{A|P^B(-x)})$ and
$p(-x)$}\\
\hline
($z_3,z_2,z_1$)& $p(-x)$& $\lambda^\prime_{\pm}$\\
\hline
(1,0,0) & $\frac{1+b_3\tanh x}{2}$ &$\frac{1}{2}\scriptstyle \left[1\pm\frac{a_3+c_3\tanh x}{1+b_3\tanh x}\right]$ \\
\hline
(0,0,1)& $\frac{1}{2}$& $\frac{1}{2}\scriptstyle \left[1\pm|a_3|\cdot\tanh x \right]$\\
\hline
(0,1,0)& $\frac{1}{2}$& $\scriptstyle\frac{1}{2} \left[1\pm\sqrt{a_3^2+(|c_2|^2-|c_1-c_2|^2-|c_1+c_2|^2+(\Re(c_1-c_2)+\Im(c_1+c_2))^2+(\Re(c_1+c_2)+\Im(c_1-c_2))^2)\tanh^2 x }\right]$\\
\hline
(0,-1,0)& $\frac{1}{2}$& ${\scriptstyle\frac{1}{2} \left[1\pm\sqrt{a_3^2+(|c_2|^2-|c_1-c_2|^2-|c_1+c_2|^2+(\Re(c_1-c_2)-\Im(c_1+c_2))^2+(\Re(c_1+c_2)-\Im(c_1-c_2))^2)\tanh^2 x }\right]}$\\
\hline
\end{tabular}
\end{align*}
From the above tables, for a given state $\rho_X$, one can get the
minimum values of $\lambda_{\pm}$ and $\lambda^\prime_{\pm}$, which
give rise to
\begin{align}
S_w(\rho_{A|P^B(+x)})=-\lambda_{+}\log_2 \lambda_{+}-\lambda_{-}\log_2 \lambda_{-},\\
S_w(\rho_{A|P^B(-x)})=-\lambda_{+}^{\prime}\log_2 \lambda_{+}^{\prime}-\lambda_{-}^{\prime}\log_2 \lambda_{-}^{\prime},
\end{align}
and the super quantum discord
\begin{align}
D_w(\rho_X)=p(+x)S_w(\rho_{A|P^B(+x)})+p(-x)S_w(\rho_{A|P^B(-x)})+S(\rho_X^B)-S(\rho_X).
\end{align}

\section{DYNAMICS OF SUPER QUANTUM DISCORD UNDER NONDISSIPATIVE CHANNELS}

In this section, firstly we will verify our formulae with  examples and illustrate that it is an extension of results in [22-23]. The first one is Werner state \cite{Werner} which is known to be a special X-state,
\begin{align*}
\rho_W=\left(\begin{array}{cccc}\frac{1+z}{4} & 0 & 0 &
\frac{z}{2}\\0& \frac{1-z}{4} & 0 & 0 \\0 &0 & \frac{1-z}{4} &
0\\\frac{z}{2} & 0 & 0 & \frac{1+z}{4}\end{array}\right).
\end{align*}
Based on formulae of the previous section, we are able to calculate eigenvalues $\lambda_{\pm i}=\lambda^{\prime}_{\pm i}=\frac{{1\pm z\tanh x}}{2}$, $i=1,2,3,4$. As everyone knows, for the Werner state, all eigenvalues get the same results for any measurement basis. The eigenvalues of $\rho_W^B$ are $\frac{1}{2},\frac{1}{2}$, and the eigenvalues of $\rho_W$ are $\frac{1+3z}{4},\frac{1-z}{4},\frac{1-z}{4},\frac{1-z}{4}$. From Eq.(16) the super quantum discord of $\rho_W$ is
$$D_{w}=-\frac{1-z\tanh x}{2}\log_{2}\frac{1-z\tanh x}{2}-\frac{1+z\tanh x}{2}\log_{2}\frac{1+z\tanh x}{2}+1+\frac{3(1-z)}{4}\log_{2}\frac{1-z}{4}+\frac{1+3z}{4}\log_{2}\frac{1+3z}{4},$$
which is in coincident with the result in \cite{Pati}.

As another example, we consider the Bell diagonal states \cite{DiVincenzo}
\begin{align*}
\rho=\left(\begin{array}{cccc}\frac{1+c_3}{4} & 0 & 0 & \frac{c_1-c_2}{4}\\0& \frac{1-c_3}{4} & \frac{c_1+c_2}{4} & 0 \\
0 &\frac{c_1+c_2}{4} & \frac{1-c_3}{4} & 0\\\frac{c_1-c_2}{4} & 0 & 0 & \frac{1+c_3}{4}\end{array}\right).
\end{align*}
From Eqs.(10)-(13) we get $\lambda_{\pm 1}=\lambda^{\prime}_{\pm 1}=\frac{1\pm c_1\tanh x}{2}$, $\lambda_{\pm 2}=\lambda^{\prime}_{\pm 2}=\frac{1\pm c_2\tanh x}{2}$, $\lambda_{\pm 3}=\lambda^{\prime}_{\pm 3}=\frac{1\pm c_3\tanh x}{2}$, $\lambda_{\pm 4}=\lambda^{\prime}_{\pm 4}=\frac{1\pm c_3\tanh x}{2}$. It is also easy to calculate the eigenvalues of $\rho^B$ are $\frac{1}{2},\frac{1}{2}$ and the eigenvalues of $\rho$ are $\frac{1-c_1-c_2-c_3}{4},\frac{1-c_1+c_2+c_3}{4},\frac{1+c_1-c_2+c_3}{4},\frac{1+c_1+c_2-c_3}{4}$.
Let $c=\max \{c_1,c_2,c_3\}$, by Eq.(16), we have the super quantum discord
\begin{eqnarray}
D_w=&-&\frac{1-c\tanh x}{2}\log_2\frac{1-c\tanh x}{2}-\frac{1+c\tanh x}{2}\log_2\frac{1+c\tanh x}{2}+1+\frac{1-c_1-c_2-c_3}{4}\log_2\frac{1-c_1-c_2-c_3}{4}\nonumber\\
&+&\frac{1-c_1+c_2+c_3}{4}\log_2\frac{1-c_1+c_2+c_3}{4}
+\frac{1+c_1-c_2+c_3}{4}\log_2\frac{1+c_1-c_2+c_3}{4}\nonumber\\
 &+&\frac{1+c_1+c_2-c_3}{4}\log_2\frac{1+c_1+c_2-c_3}{4},\nonumber
\end{eqnarray}
which coincides with the result in \cite{Wang}.

By above examples we illustrated how to apply the main result and recover the results in Refs.[22-23] as special cases.

Due to the fundamental significance and potential applications of super quantum discord, the evolution of super quantum discord under bit-flip noise  which characterized by Kraus operators 
\begin{align}
E_0=\sqrt{p}\left(\begin{array}{cc}1& 0 \\0& 1 \end{array}\right),
~~E_1=\sqrt{1-p}\left(\begin{array}{cc}0& 1 \\1& 0
\end{array}\right)
\end{align}
has been considered.
We have the channel ``local bit-flip($\Lambda_{\rm lbf}$)" :
\begin{equation}
\Lambda_{\rm lbf}(\rho_{X})=(I\otimes E_0)\rho_X(I\otimes E_0)^\dagger+(I\otimes E_1)\rho_X(I\otimes E_1)^\dagger.\\
\end{equation}
Under this channel, the entries of the density matrix have the
following transformations:
\begin{align*}
\begin{tabular}{|l|c|c|c|}
\hline
channel& $a_{11}$& $a_{14}$ & $a_{22}$\\
\hline
local bit-flip & $a_{22} + p a_{11} - p a_{22}$ & $a_{23} + p a_{14} - p a_{23}$ & $a_{11} - pa_{11} + pa_{22}$ \\
\hline
channel& $a_{23}$ &$a_{33}$ & $a_{44}$\\
\hline
local bit-flip & $a_{14} - p a_{14} + p a_{23}$ &$a_{44} + p a_{33} - p a_{44}$ & $a_{33} - p a_{33} + p a_{44}$\\
\hline
channel& $a_{23}^*$& $a_{14}^*$&\\
\hline
local bit-flip & $a_{14}^* - p a_{14}^* + p a_{23}^*$ & $a_{23}^* + p a_{14}^* - p a_{23}^*$&\\
\hline
\end{tabular}
\end{align*}

As an illustrative example, we choose a subfamily of X-types. Let us consider
\begin{align}\label{exam}
\rho_{X}=\left(\begin{array}{cccc}0.25 & 0 & 0 & 0.0625\\0& 0.25 &
0.125 & 0 \\0 & 0.125 & 0.25 & 0\\0.0625 & 0 & 0 &
0.25\end{array}\right).
\end{align}
Since quantum discord has also been employed in the study of quantum computation as an important resource, we prefer to compare super quantum discord and discord under noisy channels. It can be seen from Fig.1 that the super quantum discord attains the maximum value at $x=0$, where the weak measurement is the weakest. When $x\rightarrow \infty$, the super quantum discord approaches quantum discord.

\begin{figure}[H]
\centering
\includegraphics[width=0.5\textwidth]{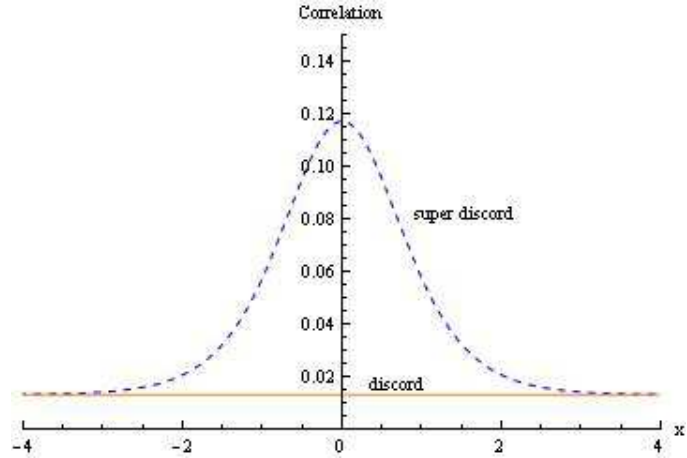}
\caption{Super quantum discord (dashed line) and quantum discord (solid line) as function of $x$.}
\end{figure}

Interestingly, the above relation motives us to introduce the super quantum discord and discord when the signal through noisy channels. From Eqs.(14)-(16), we get $p(+x)=p(-x)=0.5$,
$\lambda_{+}=\lambda_{+}^{\prime}=\max\{0.5, ~0.5+0.1875\cdot\tanh
x, ~0.5+0.0625\cdot\tanh x\}$,
$\lambda_{-}=\lambda_{-}^{\prime}=\min\{0.5, ~0.5-0.1875\cdot\tanh
x, ~0.5-0.0625\cdot\tanh x\}$. Due to the symmetry of $\tanh x$, we
take $x>0$. Namely, $p(+x)=p(-x)=0.5$,
$\lambda_{+}=\lambda_{+}^{\prime}=0.5+0.1875\cdot\tanh x$,
$\lambda_{-}=\lambda_{-}^{\prime}=0.5-0.1875\cdot\tanh x$.
Under the local bit-flip channel, we have $p(+x)=p(-x)=0.5$,
$\lambda_{+}=\lambda_{+}^{\prime}=0.5+0.1875\cdot\tanh x$,
$\lambda_{-}=\lambda_{-}^{\prime}=0.5-0.1875\cdot\tanh x$.

\begin{figure}[H]
\centering
\includegraphics[width=0.5\textwidth]{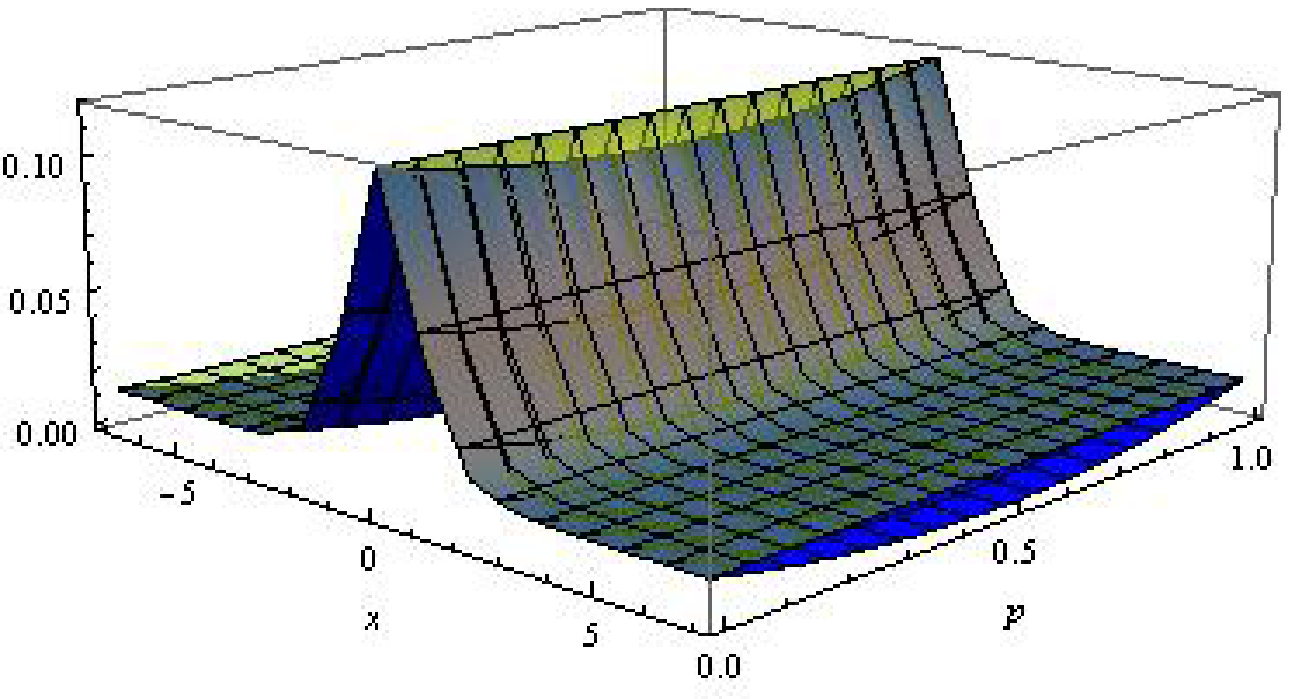}~~~~\includegraphics[width=0.5\textwidth]{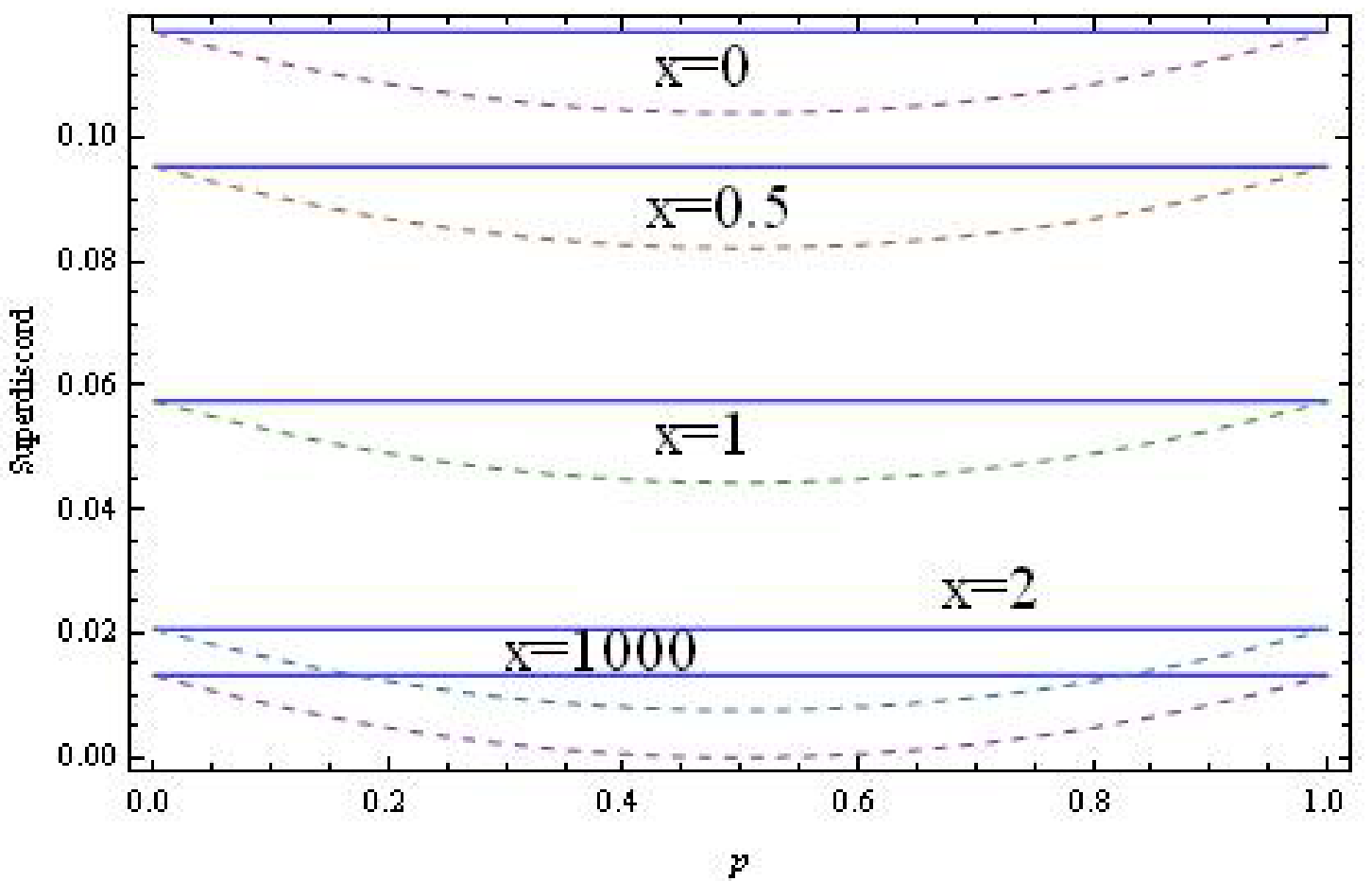}
\caption{Super quantum discord (dotted line) and super quantum discord (solid line) under local bit-flip channel of (\ref{exam}).}
\end{figure}

On the left side of Fig.2 we find that the supper quantum discord which not through the bit-flip noise channel is larger than the supper quantum discord which through the bit-flip noise channel. We can get more detail properties from the right side of Fig.2. The five curves from top to bottom are obtained by choosing the controlling parameters as the weak measurement parameter $x=0, 0.5, 1, 2,$ and $1000$. It can be seen that at $p=0$ or $p=1$, the super quantum discord is invariant under local bit-flip channel.

From a practical point of view, we sent signals through the bit-flip channel which leaves the qubit untouched with probability $p$, and flips the qubits with probability $1-p$. So through bit-flip channel the state $|0\rangle$ was taken to $|1\rangle$ for $p=0$. On the contrary, for $p=1$ the state keeps invariant. In this view, the state $\rho_x$ was not taken $|0\rangle$ to $|1\rangle$, the super quantum discord also did not change when $p=0$ or $p=1$.

In view of above argument, we then conclude that the super quantum discord will decay after through noisy channels. It means that we will lose information after the signal through noisy, hence have to control the noise probability. Furthermore, the affection of local bit-flip channel for the super quantum discord is symmetric and attains the minimum at $p=0.5$, so the noise probability plays a symmetric role in this noisy channel. It will disappear when the noise probability attaints half of one.

When considering quantum correlations captured by the super quantum discord, it is usually known that the weak measurement captures more information than POVM measurement. However, there are some counterintuitive phenomena in our example when we compare the super quantum discord after noisy channel with discord after noisy channel.

\begin{figure}[H]
\centering
\includegraphics[width=0.6\textwidth]{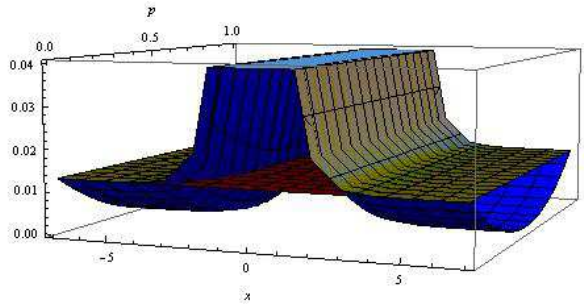}
\caption{Quantum discord (red plane) and super quantum discord (blue surface)}
\end{figure}

On the Fig.3 we can see that the red plane through the blue curve, some blue surface above red plane and others under red plane. That means after local bit-flip channel, super quantum discord is smaller than the normal quantum discord, and this difference is most obvious when the noise probability $p$ near $0.5$. Although the super quantum discord is smaller than the quantum discord only in the region of $0.01$ to $0.02$, it shows in this example, the weak measurement does not capture more information than POVM measurement. Thus, super quantum discord is a different resource than quantum discord.

\section{Conclusions and discussions}
Super quantum discord is a fundamental resource in quantum information. We have studied it for X-type states which including the Werner states and Bell diagonal states. Explicit formulae of super quantum discord for X-type states have been derived. The evolution of these states under local bit-flip channel has been investigated and reveal more different properties between super quantum discord and quantum discord. The relations between the super quantum discord and discord, evolution of super quantum discord and the week measurement strength have been analyzed.

\bigskip
\noindent{\sf Acknowledgement}
The work is supported by NSFC11275131, NSFC11305105 and KZ201410028033.


\begin{thebibliography}{9}
\bibitem{nielsen} M.A. Nielsen and I.L. Chuang, \emph{Quantum Computation and Quantum Information}
          (Cambridge University Press, Cambridge, UK, 2000).

\bibitem{bennett} C.H. Bennett, D.P. DiVincenzo, C.A. Fuchs, T. Mor, E. Rains, P.W. Shor,
             J.A. Smolin, and W.K. Wootters, Phys. Rev. A \textbf{59}, 1070 (1999).

\bibitem{horodecki2} M. Horodecki, P. Horodecki,  R. Horodecki, J. Oppenheim, A. Sen, U. Sen,
              and B. Synak-Radke,  Phys. Rev. A \textbf{71}, 062307 (2005).

\bibitem{niset} J. Niset and N.J. Cerf, Phys. Rev. A \textbf{74}, 052103 (2006).

\bibitem{ollivier} H. Ollivier and W.H. Zurek, Phys. Rev. Lett. \textbf{88}, 017901 (2001).

\bibitem{bylicka} B. Bylicka and D. Chru\'{s}ci\'{n}ski, Phys. Rev. A \textbf{81}, 062102 (2010).

\bibitem{werlang} T. Werlang, S. Souza, F.F. Fanchini, and C.J. Villas Boas, Phys. Rev. A \textbf{80}, 024103 (2009).

\bibitem{sarandy} M.S. Sarandy, Phys. Rev. A \textbf{80}, 022108 (2009).

\bibitem{ferraro} A.~Ferraro, L.~Aolita, D.~Cavalcanti, F.~M. Cucchietti, and
A.~Ac{\'\i}n, Phys. Rev. A \textbf{81}, 052318 (2010).

\bibitem{fanchini} F.F. Fanchini, T. Werlang, C.A. Brasil, L.G.E. Arruda, and A.O. Caldeira, Phys. Rev. A \textbf{81}, 052107 (2010).

\bibitem{dakic} B.~Dak{\'\i}c, V.~Vedral, and {\v C}.~Brukner, Phys. Rev. Lett. \textbf{105}, 190502 (2010).

\bibitem{modi} K. Modi, T. Paterek, W. Son, V. Vedral, and M. Williamson, Phys. Rev. Lett. \textbf{104}, 080501
(2010).

\bibitem{luo1} N. Li and S. Luo, Phys. Rev. A \textbf{76}, 032327 (2007); S. Luo, {\it ibid} {\bf 77}, 022301 (2008).

\bibitem{luo} S. Luo, Phys. Rev. A \textbf{77}, 042303 (2008).

\bibitem{libo} B. Li, Z.X. Wang and S.M. Fei, Phys. Rev. A {\bf 83},  022321(2011).

\bibitem{lang} M.D. Lang, and C.M. Caves, Phys. Rev. Lett. \textbf{105}, 150501 (2010).

\bibitem{ali}  M. Ali, A.R.P. Rau, and G. Alber, Phys. Rev. A \textbf{81}, 042105 (2010);
M. Ali, A.R.P. Rau, and G. Alber, {\it ibid} {\bf 82}, 069902
(2010).

\bibitem{Mazzola} L. Mazzola, J. Piilo, and S. Maniscalco, Phys. Rev. Lett. \textbf{104}, 200401 (2010).

\bibitem{Maziero} J.~Maziero, L.~C. C{\'e}leri, R.~M. Serra, and V.~Vedral, Phys. Rev. A \textbf{80}, 044102 (2009).

\bibitem{ost}L. Roa, J. C. Retamal, M. Alid-Vaccarezza, Phys. Rev. Lett. \textbf{107}, 080401 (2011);
B. Li, S.M. Fei, Z.X. Wang and H. Fan, Phys. Rev. A \textbf{85},
022328 (2012).

\bibitem{Aharonov}Y. Aharonov, D.Z. Albert, and L. Vaidman, Phys. Rev. Lett. \textbf{60}, 1351 (1998).

\bibitem{Pati}U. Singh and A.K. Pati, Annals of Physics 343,141(2014).

\bibitem{Wang}Y.K. Wang, T. Ma, H. Fan, S.M. Fei, and Z.X. Wang, Quantum Inf. Process. \textbf{13}, 283 (2014).

\bibitem{Shi}M. Shi, C. Sun, F. Jiang, X. Yan and J. Du, Phys. Rev. A  \textbf{85}, 064104 (2012).

\bibitem{Partovi}M. H. Partovi, Phys. Lett. A \textbf{137}, 455 (1989).

\bibitem{Henderson}L. Henderson and V. Vedral, J. Phys. A \textbf{34}, 6899 (2001).

\bibitem{Oreshkov}O. Oreshkov and T. A. Brun, Phys. Rev. Lett. \textbf{95}, 110409 (2005).


\bibitem{Werner}R.F. Werner, Phys. Rev. A \textbf{40}, 4277 (1989).

\bibitem{DiVincenzo}C.H. Bennett, D.P. DiVincenzo, J.A. Smolin, and W.K. Wootters, Phys. Rev. A \textbf{54}, 3824 (1996).

\end{thebibliography}
\end{document}